%% file: main.tex
\documentclass[runningheads]{llncs}

\usepackage[T1]{fontenc}
\usepackage[utf8]{inputenc}
\usepackage{amsmath,amssymb}
\usepackage{mathtools}

\usepackage[dvipsnames]{xcolor}

\usepackage{enumitem}

\usepackage{url}
\usepackage{xurl} 
\usepackage{hyperref}
\usepackage[nameinlink,capitalize,noabbrev]{cleveref} 

\usepackage{booktabs}
\usepackage{caption}
\usepackage{threeparttable}
\usepackage{rotating}

\hypersetup{
  colorlinks=true,
  linkcolor=Blue,
  citecolor=Blue,
  urlcolor=Blue,
}

\spnewtheorem{assump}{Assumption}{\bfseries}{\itshape}
\spnewtheorem{lem}{Lemma}{\bfseries}{\itshape}
\spnewtheorem{prop}{Proposition}{\bfseries}{\itshape}
\spnewtheorem{cor}{Corollary}{\bfseries}{\itshape}
\spnewtheorem{hypo}{Hypothesis}{\bfseries}{\itshape}
\spnewtheorem{defn}{Definition}{\bfseries}{\itshape}

\newcommand{\E}{\mathbb{E}}
\newcommand{\Var}{\mathrm{Var}}
\newcommand{\CE}{\mathrm{CE}}
\newcommand{\IV}{\mathrm{IV}}

\newcommand{\RV}{\mathrm{RV}}

\newcommand{\bigO}{\mathcal{O}}

\title{Impact of Volatility on Time-Based Transaction Ordering Policies}
\titlerunning{Volatility and Time-Based Ordering Policies}

\author{Ko Sunghun\inst{1,2} \and Jinsuk Park\inst{1}}
\authorrunning{K.~Sunghun and J.~Park}

\institute{
  Matroos Labs \and
  Department of Mathematical Sciences, KAIST, Daejeon, Republic of Korea\\
  \email{\{kosunghun317,jinsuk\}@matroos.xyz}
}

\begin{document}
\maketitle

\begin{abstract}
We study Arbitrum’s \emph{Express Lane Auction} (ELA), an ahead-of-time second-price auction that grants the winner an exclusive latency advantage for one minute. Building on a single-round model with risk-averse bidders, we propose a hypothesis that the value of priority access is discounted relative to risk-neutral valuation due to the difficulty of forecasting short-horizon volatility and bidders' risk aversion. We test these predictions using ELA bid records matched to high-frequency ETH prices and find that the result is consistent with the model.
\keywords{Decentralized finance \and MEV \and Transaction ordering}
\end{abstract}

\input{Introduction}
\input{Theory}
\input{Empirical_Analysis}

\input{Conclusion}

\bibliographystyle{splncs04}
\bibliography{refs}

\appendix
\section{Tables}

\begin{table}[htbp]
\centering
\begin{tabular}{lcccc}
\toprule
 & \multicolumn{2}{c}{Bidder 0x8c6f} & \multicolumn{2}{c}{Bidder 0x95c0} \\
\cmidrule(lr){2-3} \cmidrule(lr){4-5}
Variable & Reduced & Full & Reduced & Full \\
\midrule
\multicolumn{5}{l}{\textit{Panel A: Location Coefficients}} \\
\addlinespace[0.5em]
Intercept ($\theta_0$) & 44.5019$^{***}$ & 44.7182$^{***}$ & 45.4140$^{***}$ & 45.7480$^{***}$ \\
 & (0.0313) & (0.0298) & (0.0283) & (0.0275) \\
$\log$(E[IV]/$\sqrt{P}$) ($\theta_1$) & 0.4541$^{***}$ & 1.2026$^{***}$ & 0.4962$^{***}$ & 1.3663$^{***}$ \\
 & (0.0016) & (0.0048) & (0.0015) & (0.0044) \\
$\log$(Var(IV)) ($\theta_2$) & -- & -0.4403$^{***}$ & -- & -0.5092$^{***}$ \\
 & -- & (0.0026) & -- & (0.0024) \\
\addlinespace[0.5em]
\midrule
\multicolumn{5}{l}{\textit{Panel B: Scale Coefficients}} \\
\addlinespace[0.5em]
Intercept ($\gamma_0$) & -0.0320 & 0.5389$^{***}$ & -0.5813$^{***}$ & -0.2515$^{***}$ \\
 & (0.0169) & (0.0206) & (0.0162) & (0.0204) \\
$\log$(E[IV]/$\sqrt{P}$) ($\gamma_1$) & -0.0013 & 0.0384$^{***}$ & -0.0255$^{***}$ & -0.0842$^{***}$ \\
 & (0.0009) & (0.0038) & (0.0008) & (0.0039) \\
$\log$(Var(IV)) ($\gamma_2$) & -- & -0.0042$^{*}$ & -- & 0.0478$^{***}$ \\
 & -- & (0.0021) & -- & (0.0021) \\
\addlinespace[0.5em]
\midrule
\midrule
\multicolumn{5}{l}{\textit{Panel C: Model Fit Statistics}} \\
\addlinespace[0.5em]
Observations & 264959 & 264959 & 264959 & 264959 \\
Censored & 9151360 & 9151360 & 9151360 & 9151360 \\
Log Likelihood & -358833.46 & -345110.74 & -338519.76 & -317455.38 \\
AIC & 717674.91 & 690233.48 & 677047.52 & 634922.76 \\
BIC & 717716.86 & 690296.41 & 677089.47 & 634985.69 \\
McFadden $R^2$ & 0.119 & 0.153 & 0.150 & 0.203 \\
\addlinespace[0.5em]
\midrule
\multicolumn{5}{l}{\textit{Panel D: Full vs Reduced Model Comparison}} \\
\addlinespace[0.5em]
$\Delta$ AIC (Full - Reduced) & \multicolumn{2}{c}{-27441.43} & \multicolumn{2}{c}{-42124.76} \\
$\Delta$ BIC (Full - Reduced) & \multicolumn{2}{c}{-27420.45} & \multicolumn{2}{c}{-42103.78} \\
LR Test $\chi^2$ (p-val) & \multicolumn{2}{c}{$27445.43^{***}$} & \multicolumn{2}{c}{$42128.76^{***}$} \\
\bottomrule
\end{tabular}
\caption{Heteroskedastic Tobit (Log-Log, Gaussian Errors). $^{*}p<0.05$, $^{**}p<0.01$, $^{***}p<0.001$.}
\label{tab:log_nolag}
\end{table}

\begin{table}[htbp]
\centering
\caption{Heteroskedastic Tobit Regression by Subsample}
\label{tab:tobit_subsample}
\footnotesize
\begin{tabular}{llccccccc}
\toprule
Bidder & Subsample & N & $\theta_1^R$ & $\theta_1^F$ & $\theta_2^F$ & LR & $\Delta$AIC & $\Delta$BIC \\
\midrule
0x8c6f & 2025-05 & 44640 & 0.070 & 0.071 & -0.080 & 571.3 & -567.3 & -549.9 \\
 & 2025-06 & 43200 & 0.425 & 0.485 & -2.713 & 1809.4 & -1805.4 & -1788.1 \\
 & 2025-07 & 44640 & 0.365 & 0.423 & -2.380 & 1929.8 & -1925.8 & -1908.4 \\
 & 2025-08 & 44640 & 0.179 & 0.182 & -1.103 & 1081.7 & -1077.7 & -1060.3 \\
 & 2025-09 & 43200 & 0.018 & 0.026 & -0.207 & 812.9 & -808.9 & -791.6 \\
 & 2025-10 & 44639 & 0.167 & 0.158 & -0.377 & 892.3 & -888.3 & -870.9 \\
\addlinespace[0.3em]
 & High E[IV]/$\sqrt{P}$ & 132479 & 0.276 & 0.329 & -1.612 & 4196.5 & -4192.5 & -4172.9 \\
 & Low E[IV]/$\sqrt{P}$ & 132480 & 0.428 & 0.599 & -86.673 & 10163.1 & -10159.1 & -10139.6 \\
\midrule
0x95c0 & 2025-05 & 44640 & 0.461 & 0.483 & -2.385 & 1681.3 & -1677.3 & -1659.9 \\
 & 2025-06 & 43200 & 0.538 & 0.581 & -3.640 & 2561.0 & -2557.0 & -2539.7 \\
 & 2025-07 & 44640 & 0.450 & 0.481 & -3.101 & 2456.7 & -2452.7 & -2435.3 \\
 & 2025-08 & 44640 & 0.228 & 0.252 & -2.330 & 1564.9 & -1560.9 & -1543.5 \\
 & 2025-09 & 43200 & 0.259 & 0.285 & -2.372 & 1005.9 & -1001.9 & -984.6 \\
 & 2025-10 & 44639 & 0.304 & 0.338 & -2.166 & 1885.6 & -1881.6 & -1864.2 \\
\addlinespace[0.3em]
 & High E[IV]/$\sqrt{P}$ & 132479 & 0.332 & 0.398 & -2.295 & 6232.3 & -6228.3 & -6208.7 \\
 & Low E[IV]/$\sqrt{P}$ & 132480 & 0.514 & 0.896 & -145.477 & 7800.4 & -7796.4 & -7776.8 \\
\bottomrule
\end{tabular}
\begin{tablenotes}
\footnotesize
\item \textit{Notes:} $\theta_1^R$: E[IV]/$\sqrt{P}$ coefficient (reduced model), $\theta_1^F$: E[IV]/$\sqrt{P}$ coefficient (full model), $\theta_2^F$: Var(IV) coefficient (full model). High/Low regimes split at the median of E[IV]/$\sqrt{P}$ across the entire dataset. All LR tests are significant at $p<0.001$.
\end{tablenotes}
\end{table}

\begin{table}[htbp]
\centering
\begin{tabular}{lcccc}
\toprule
 & \multicolumn{2}{c}{Bidder 0x8c6f} & \multicolumn{2}{c}{Bidder 0x95c0} \\
\cmidrule(lr){2-3} \cmidrule(lr){4-5}
Variable & Reduced & Full & Reduced & Full \\
\midrule
\multicolumn{5}{l}{\textit{Panel A: Location Coefficients}} \\
\addlinespace[0.5em]
Intercept ($\theta_0$) & 0.9239$^{***}$ & 0.8976$^{***}$ & 0.9783$^{***}$ & 0.9646$^{***}$ \\
 & (0.0024) & (0.0025) & (0.0019) & (0.0027) \\
$\mathbb{E}[IV]/\sqrt{P}$ ($\theta_1$) & 0.4234$^{***}$ & 0.4306$^{***}$ & 0.5134$^{***}$ & 0.5318$^{***}$ \\
 & (0.0012) & (0.0013) & (0.0011) & (0.0013) \\
$\text{Var}(IV)$ ($\theta_2$) & -- & -0.9335$^{***}$ & -- & -1.7999$^{***}$ \\
 & -- & (0.0279) & -- & (0.0364) \\
\addlinespace[0.5em]
\midrule
\multicolumn{5}{l}{\textit{Panel B: Scale Coefficients}} \\
\addlinespace[0.5em]
Intercept ($\gamma_0$) & -0.2747$^{***}$ & -1.6680$^{***}$ & 0.1052$^{***}$ & -0.8891$^{***}$ \\
 & (0.0048) & (0.0260) & (0.0042) & (0.0235) \\
$\log(\mathbb{E}[IV]/\sqrt{P})$ ($\gamma_1$) & 0.6417$^{***}$ & 0.9775$^{***}$ & 0.5029$^{***}$ & 0.7404$^{***}$ \\
 & (0.0020) & (0.0065) & (0.0017) & (0.0059) \\
$\log(\text{Var}(IV))$ ($\gamma_2$) & -- & -0.1976$^{***}$ & -- & -0.1418$^{***}$ \\
 & -- & (0.0036) & -- & (0.0032) \\
\addlinespace[0.5em]
\midrule
\midrule
\multicolumn{5}{l}{\textit{Panel C: Model Fit Statistics}} \\
\addlinespace[0.5em]
Observations & 264958 & 264958 & 264958 & 264958 \\
Censored & 264958 & 264958 & 264958 & 264958 \\
Student $t$ $\nu$ & 1.54 & 1.61 & 2.09 & 2.22 \\
Log Likelihood & -685494.19 & -683003.61 & -676359.13 & -672856.17 \\
AIC & 1370998.37 & 1366021.22 & 1352728.26 & 1345726.35 \\
BIC & 1371050.81 & 1366094.63 & 1352780.70 & 1345799.76 \\
McFadden $R^2$ & 0.128 & 0.131 & 0.139 & 0.144 \\
\addlinespace[0.5em]
\midrule
\multicolumn{5}{l}{\textit{Panel D: Full vs Reduced Model Comparison}} \\
\addlinespace[0.5em]
$\Delta$ AIC (Full - Reduced) & \multicolumn{2}{c}{-4977.15} & \multicolumn{2}{c}{-7001.92} \\
$\Delta$ BIC (Full - Reduced) & \multicolumn{2}{c}{-4956.18} & \multicolumn{2}{c}{-6980.94} \\
LR Test $\chi^2$ (p-val) & \multicolumn{2}{c}{$4981.15^{***}$} & \multicolumn{2}{c}{$7005.92^{***}$} \\
\bottomrule
\end{tabular}
\caption{Regression with Lagged IV estimators (Linear, t-dist Errors). $^{*}p<0.05$, $^{**}p<0.01$, $^{***}p<0.001$.}
\label{tab:linear_lagged}
\end{table}

\begin{table}[htbp]
\centering
\begin{tabular}{lcccc}
\toprule
 & \multicolumn{2}{c}{Bidder 0x8c6f} & \multicolumn{2}{c}{Bidder 0x95c0} \\
\cmidrule(lr){2-3} \cmidrule(lr){4-5}
Variable & Reduced & Full & Reduced & Full \\
\midrule
\multicolumn{5}{l}{\textit{Panel A: Location Coefficients}} \\
\addlinespace[0.5em]
Intercept ($\theta_0$) & 47.1042$^{***}$ & 47.0202$^{***}$ & 47.7172$^{***}$ & 47.7364$^{***}$ \\
 & (0.0285) & (0.0301) & (0.0249) & (0.0268) \\
$\log(\mathbb{E}[IV]/\sqrt{P})$ ($\theta_1$) & 0.5892$^{***}$ & 1.0764$^{***}$ & 0.6160$^{***}$ & 1.1954$^{***}$ \\
 & (0.0015) & (0.0045) & (0.0013) & (0.0041) \\
$\log(\text{Var}(IV))$ ($\theta_2$) & -- & -0.2937$^{***}$ & -- & -0.3455$^{***}$ \\
 & -- & (0.0026) & -- & (0.0023) \\
\addlinespace[0.5em]
\midrule
\multicolumn{5}{l}{\textit{Panel B: Scale Coefficients}} \\
\addlinespace[0.5em]
Intercept ($\gamma_0$) & -0.6782$^{***}$ & -0.1254$^{***}$ & -1.4440$^{***}$ & -1.1117$^{***}$ \\
 & (0.0157) & (0.0196) & (0.0146) & (0.0190) \\
$\log(\mathbb{E}[IV]/\sqrt{P})$ ($\gamma_1$) & -0.0287$^{***}$ & 0.0391$^{***}$ & -0.0637$^{***}$ & -0.1032$^{***}$ \\
 & (0.0008) & (0.0036) & (0.0008) & (0.0036) \\
$\log(\text{Var}(IV))$ ($\gamma_2$) & -- & -0.0224$^{***}$ & -- & 0.0353$^{***}$ \\
 & -- & (0.0019) & -- & (0.0019) \\
\addlinespace[0.5em]
\midrule
\midrule
\multicolumn{5}{l}{\textit{Panel C: Model Fit Statistics}} \\
\addlinespace[0.5em]
Observations & 264958 & 264958 & 264958 & 264958 \\
Censored & 9151325 & 9151325 & 9151325 & 9151325 \\
Log Likelihood & -327469.69 & -320362.18 & -305628.46 & -294324.82 \\
AIC & 654947.38 & 640736.36 & 611264.92 & 588661.64 \\
BIC & 654989.33 & 640799.28 & 611306.86 & 588724.57 \\
McFadden $R^2$ & 0.196 & 0.214 & 0.233 & 0.261 \\
\addlinespace[0.5em]
\midrule
\multicolumn{5}{l}{\textit{Panel D: Full vs Reduced Model Comparison}} \\
\addlinespace[0.5em]
$\Delta$ AIC (Full - Reduced) & \multicolumn{2}{c}{-14211.02} & \multicolumn{2}{c}{-22603.27} \\
$\Delta$ BIC (Full- Reduced) & \multicolumn{2}{c}{-14190.04} & \multicolumn{2}{c}{-22582.30} \\
LR Test $\chi^2$ (p-val) & \multicolumn{2}{c}{$14215.02^{***}$} & \multicolumn{2}{c}{$22607.27^{***}$} \\
\bottomrule
\end{tabular}
\caption{Regression with Lagged IV estimators (Log-Log, Gaussian Errors). $^{*}p<0.05$, $^{**}p<0.01$, $^{***}p<0.001$.}
\label{tab:log_lagged}
\end{table}

\end{document}

%% file: Introduction.tex
\section{Introduction}
\subsection{Background}
Timeboost, a novel transaction ordering policy introduced in \cite{mamageishvili2023buying} and subsequently implemented by Offchain Labs, has recently been deployed on the Arbitrum mainnet \cite{offchainlabs2025timeboost}. Under the Timeboost policy, transactions are delivered to the sequencer through two distinct channels: the normal lane and the Express Lane (EL). While the sequencer processes transactions in a First-In, First-Out (FIFO) manner, transactions submitted via the normal lane incur a $200$-millisecond delay before being forwarded to the sequencer. Conversely, the EL immediately forwards received transactions, thereby offering its users a $200$-millisecond latency advantage.

The temporary right to utilize the EL is periodically allocated for each round—which lasts for one minute—via a second-price sealed-bid auction termed the \emph{Express Lane Auction (ELA)}\footnote{The winner may elect to either exclusively utilize this access or to share and resell it to other parties. For instance, \cite{kairossubauction} is currently running a secondary market that resells the EL access.}. This auction operates on an ahead-of-time basis; participants in the ELA are required to estimate the value of the good (the right to use the EL in the forthcoming round), submit their bids, and pay if they win, prior to exercising their EL access. Given that the latency differential is barely noticeable to typical users, participation and bidding in the ELA are primarily confined to latency-sensitive players, such as High-Frequency Traders (HFTs). The primary objective of this mechanism is to capture a portion of the Maximal Extractable Value (MEV) generated, without compromising the experience of the typical user, by imposing charges solely on strategic players, frequently referred to as \emph{searchers}. 

Searchers value access to EL due to the 200-millisecond latency advantage, which is typically sufficient to secure the majority of CEX-DEX arbitrage opportunities for each arbitrage oppportunities\footnote{This does not universally hold true. For instance, if the latency advantage is insufficiently large, a probabilistic strategy involving transaction spamming can still capture a non-negligible portion of opportunities. For further details, refer to \cite{akaki2025blind}. Generally, a $200$-millisecond gap is considered big enough to mitigate such possibilities.}. Given that CEX-DEX arbitrage profit is directly proportional to the integrated variance $IV (\equiv \int^T_0 \sigma^2_t dt)$ of risky assets, bidding in ELA yields a payoff analogous to a long position in a variance swap. Previous research has identified a discount in variance swap markets, termed the \emph{Variance Risk Premium (VRP)}, which represents the discrepancy between the theoretical fair price (derived from a replicating portfolio) and the observed market price. Studies such as \cite{bollerslev2009expected} and \cite{carr2009variance} attribute the existence of VRP to the stochastic nature of volatility and traders' risk aversion. Consequently, a similar question arises for ELA: do bidders submit bids below the expected CEX-DEX arbitrage profit for an upcoming round? This study addresses questions concerning the factors influencing Timeboost's performance and their impact, including the aforementioned phenomenon.

More specifically, we endeavor to answer the following research questions: \emph{(i)} Does a discount analogous to VRP exist in ELA? \emph{(ii)} If such a discount exists, what is its primary causal mechanism? and \emph{(iii)} What is the intensity of this discount effect?

Our findings indicate that \emph{(i)} a comparable discount does exist, and \emph{(ii)} it is consistent with the inherent difficulty in forecasting the future volatility of risky assets within a short time horizon. We also provide an estimation of the intensity of this effect.

This study is one of the first empirical investigations into the performance of Timeboost, examining the determinants of bidder valuations and the efficacy of express lane auctions. As such, it establishes a foundational reference for subsequent research that employs more sophisticated methodologies and realistic assumptions. Ultimately, this investigation seeks to enhance the understanding of time-based sequencing mechanisms, including Timeboost, within auction frameworks.

\begin{figure}[htbp]
    \centering
    \includegraphics[width=0.8\linewidth]{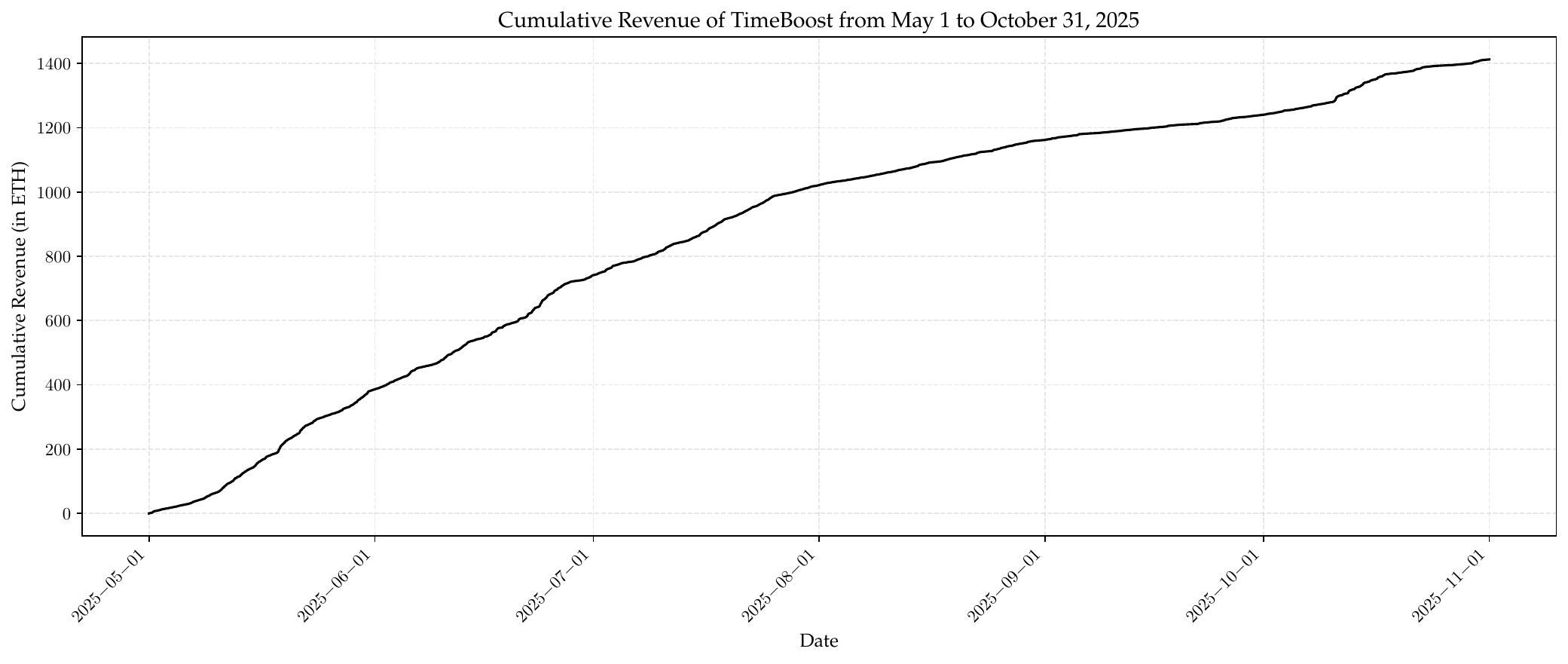}
    \caption{Cumulative revenue of Timeboost through express lane auction from May to October.}
    \label{fig:cumulative_price}
\end{figure}

\subsection{Related Literature}
\paragraph*{Loss-versus-rebalancing and CEX-DEX Arbitrage.}
Following the foundational work of \cite{daian2019flash}, which introduced the term Maximal Extractable Value (MEV), a substantial body of literature has emerged exploring both its theoretical and empirical aspects. Early on, both academia and practitioners recognized CEX-DEX arbitrage as the primary source of MEV, commonly measured using short-term ($\le 5$ minutes) markouts \cite{wu2025measuring,gogol2024cross,heimbach2024non,oz2025pandora,followupuniv3,usageofmarkout}. Subsequently, \cite{milionis2022automated} provided a theoretical justification for such a metric and introduced Loss-Versus-Rebalancing (LVR), which has been generalized to more realistic settings in later works, including \cite{milionis2024automated} and \cite{nezlobin2025loss}. These studies, while differing in their specific details, consistently demonstrate a close relationship between CEX-DEX arbitrage and LVR, which represents the adverse selection cost borne by Automated Market Makers (AMMs). After abstracting away fees, gas costs, and inventory risk, the leading term of this relationship is shown to be $\bigO (\sigma^2)$, where $\sigma$ denotes the instantaneous volatility of the asset price.

\paragraph*{Time-based Transaction Ordering Policies and Timeboost.}
Even before the advent of blockchain technology, time-based transaction ordering policies and associated products were well established and widely adopted. For instance, since the late $2000$s, traditional finance (TradFi) exchanges have commercialized latency, offering ``speed technology'' to high-frequency trading (HFT) firms, generating revenues exceeding \$ $700$ million in $2017$ \cite{budish2024theory}. The impact of asymmetric speed bumps and micro-burst fees, exemplifying such exotic transaction ordering policies (priority rules), has been investigated by \cite{Aoyagi_2025} and \cite{brolley2023liquid}. Timeboost, as introduced in \cite{mamageishvili2023buying}, aligns with these policies, and \cite{akaki2025blind} and \cite{fritsch2024mev} provide a comprehensive theoretical understanding of its mechanisms. While Timeboost generated over $1400$ ETH in revenue from May to October, demonstrating its effectiveness, empirical research in this area remains nascent, with \cite{messias2025express} and \cite{mamageishvili2025timeboostaheadoftimeauctionswork} among the pioneering works that explore bidder behavior in ELA and the performance of Timeboost. In particular, the latter work claimed that the bidders' bids can be used as a noisy proxy for the markout of arbitrageurs' trades in the subsequent round.

\paragraph*{Variance Risk Premia and Risk Aversion.}
A large body of literature documents a systematic wedge between the replication-based (risk-neutral) fair value of a variance swap and its traded price, commonly referred to as the variance (or volatility) risk premium (VRP). The ``fair'' strike for variance swap can be synthesized from an option strip \cite{demeterfi1999guide} or via model‐free implied variance \cite{jiang2005modelfree}, and the VRP is typically measured as implied minus subsequently realized variance ($\RV$) \cite{carr2009variance}. Empirically, this wedge is time-varying and (on average) negative, consistent with investors’ risk aversion and a stochastic nature of volatility that loads on variance and tail risks \cite{bakshi2003delta,bollerslev2009expected,broadie2007model,bollerslev2011tails}. Beyond level effects, the term structure and composition of premia matter as well: short-horizon variance premia are especially volatile and command higher compensation, and downside (``bad'') variance carries a larger premium than upside variance \cite{dewbecker2017price,aitsahalia2020term,feunou2018downside}. The phenomenon is not equity‐specific: analogous variance premia appear in foreign exchange (FX) and commodity markets, indicating a broad, asset-agnostic compensation for variance and jump risks \cite{londono2017fxvrp,prokopczuk2017commodityvrp,vantassel2018equityvrp}. Collectively, these results suggest that forward-looking prices for variance-linked payoffs can be persistently discounted relative to their ``fair'' values when risk is borne by risk-averse traders.

\subsection{Roadmap} 
The subsequent sections of this paper are structured as follows: Section 2 provides a theoretical foundation for the proposed model and hypothesis. Section 3 outlines the data sources and the analytical methodologies employed, presents results, and discusses the findings and their implications. The paper concludes with Section 4, which summarizes the key findings and potential future research topics.

%% file: Theory.tex
\section{Theory}\label{sec:theory}
In this section, we provide a theoretical derivation of how bidders value the express lane (EL). We model each round of the express lane auction (ELA) as a second-price sealed-bid auction for a single, exclusive right to use the EL for a one-minute period, within a conditionally independent private value (CIPV) framework. We assume the winner captures the arbitrage opportunity solely during this period.

We first review how CEX-DEX arbitrage profit is related to the volatility of risky assets. We then specify the auction environment and timing and derive the bidders' payoffs. From the payoff, we infer bidders' valuations and their bids, and finally present a reduced-form model and hypotheses for empirical testing.

\subsection{Arbitrageurs' Profit}
Since the value of the express lane is closely related to MEV, and most of that MEV comes from CEX-DEX arbitrage, it is important to understand arbitrageurs' profits to correctly value the express lane. In this section, as mentioned earlier, we review the theory of CEX-DEX arbitrage. For simplicity, consider a single constant product market maker (CPMM) pool with an invariant \(L^2\) that consists of a risky asset and a stablecoin. We assume the log price of the risky asset at CEX, which is a \emph{true} price, follows a semi-martingale process with stochastic volatility, \[
    d\left(\log P_t \right)= \mu_t dt + \sigma_t dW_t \, .
\] \cite{milionis2022automated} then showed that loss-versus-rebalancing, which is the arbitrageur's revenue under ideal settings (i.e., no gas fee and no friction in trading), is proportional to the pool value and volatility of the risky asset.
\begin{prop}[\cite{milionis2022automated}]
    The instantaneous rate of loss-versus-rebalancing \(\overline{\mathrm{LVR}}\) is \[
    \overline{\mathrm{LVR}} = \frac{L\sqrt{P_t}}{4} \sigma^2_t,
    \] where \(P_t\) and \(\sigma_t\) are price and volatility of risky asset at time \(t\), respectively.
\end{prop}
Thus, for interval \([0,T]\), the arbitrageur's revenue \(\mathrm{ARB}\)is
\begin{align}
    \mathrm{ARB} &= \int^T_0 \overline{\mathrm{LVR}} \, dt \\
        &=  \int^T_0 \frac{L\sqrt{P_t}}{4} \sigma^2_t \, dt \\
        &\approx \frac{L\sqrt{P_0}}{4} \int^T_0 \sigma^2_t \, dt \\
        &= \frac{L\sqrt{P_0}}{4} \IV,
\end{align}
where integrated variance \(\IV \equiv \int^T_0 \sigma^2_t \, dt \), with assumption that \(T\) is small enough so that \(P_0 \approx P_t \) for all \(t \in [0,T]\). In practice, a block is not generated continuously, and the arbitrageur has to bear the gas fees and trading costs. However, as shown in works including \cite{milionis2024automated} and \cite{nezlobin2025loss}, it remains true that an arbitrageur's revenue is proportional to the pool's value and the square of the asset's volatility.

\subsection{Express Lane Auction}
Here, we clarify the mechanism of the express lane auction (ELA) and its timing. Participants in ELA compete in each round, lasting a minute, for the exclusive right to use the express lane. They want such a right because those with access to the express lane can win all the CEX-DEX arbitrage opportunities. The auction is conducted as a sealed-bid second-price auction, with the bidding deadline strictly before the start of each round. Therefore, the bidders should calculate the value of the express lane based on both public and private information provided to each bidder, and submit a bid before the value is realized. We begin with the following definitions.
\begin{defn}[Rounds, Periods, Bidders, Information, and Reward]
    Let: \begin{itemize}
        \item \(r \in \mathbb N\) be each round of ELA,
        \item \([T_r, T_{r+1})\) be period of each round,
        \item \(D_r < T_r\) be the deadline of bid submission for each round,
        \item \(S \equiv \{1, 2, \cdots, N\}\) be set of bidders,
        \item \(\mathcal F_r\) and \(\mathcal S_{ir}\) be public and private information for bidder \(i\) available at \(D_r\), respectively,
        \item \(v_{ir} \) be bidder \(i\)'s valuation at round \(r\), and
        \item \(b_{ir}\) be bidder \(i\)'s bid at round \(r\).
    \end{itemize}
\end{defn}

\subsection{Bidders' Payoff and Valuation}

We now derive how bidders value the right to use the express lane in a given
round. For notational convenience, write
\[
  \IV_r \equiv \IV_{[T_r, T_{r+1})}
  = \int_{T_r}^{T_{r+1}} \sigma_t^2 \, dt
\]
for the integrated variance over round $r$. The winner at round \(r\) will earn profit from arbitrage opportunities within that round. Thus the reward \(R_r\) is \[
R_r = \int^{T_{r+1}}_{T_r} \overline{\mathrm{LVR}} \, dt \approx \frac{L\sqrt{P_{T_r}}}{4} \IV_r,
\] where \(L\) is liquidity available and \(\IV_{[T_r, T_{r+1})}\) is integrated variance of risky asset from \(T_r\) to \(T_{r + 1}\).\footnote{While traders use the entire liquidity available, beyond a single CPMM pool, if the price does not move much within a round, we can still locally approximate it as a single big CPMM pool; see \cite{Dan2021UniswapV3Universal}. Through the paper, we assume that \(L\) is constant. It is roughly true, since the majority of liquidity is provided by professional LPs who concentrate liquidity near the current price most of the time.} Note that \(R_r\) is random variable and only fully realized after time \(T_{r+1}\). In practice, even with express lane access, a bidder may only capture a fraction of the available arbitrage, and may face idiosyncratic costs or benefits (e.g., private order flow, infrastructure frictions).\footnote{Note that this does not necessarily imply that others can take advantage of it and earn profit. If such friction exists, it is very likely that others will have to bear that friction as well, which makes trade unprofitable.} We model the \emph{net} profit of bidder $i$ from winning round $r$ as
\begin{equation}\label{eq:profit-def}
  \Pi_{ir}
  = \alpha_{ir} + \beta_{ir} \IV_r \sqrt{P_{T_r}},
\end{equation}
where:
\begin{itemize}
  \item $\alpha_{ir}$ collects components of profit and cost that do not scale with volatility (e.g., fixed costs, long-tail MEV),
  \item $\beta_{ir}$ is bidder $i$'s \emph{extraction efficiency} in round $r$, summarizing how effectively the bidder converts volatility into arbitrage revenue. In particular, $\beta_{ir}$ subsumes both the liquidity factor $L/4$ and the bidder's share of the total arbitrage $R_r$.
\end{itemize}
Both $\alpha_{ir}$ and $\beta_{ir}$ are measurable with respect to bidder $i$'s private information $\mathcal S_{ir}$ and the public information $\mathcal F_r$ available at the bidding deadline $D_r$.

Conditional on $(\mathcal F_r, \mathcal S_{ir})$, bidder $i$ forms a belief on the distribution over $\IV_r$ and thus over $\Pi_{ir}$. Let
\begin{equation}
m^{\IV}_{ir} \equiv \E\left[\IV_r \mid \mathcal F_r, \mathcal S_{ir} \right] \quad \text{and} \quad v^{\IV}_{ir} \equiv \Var\left(\IV_r \mid \mathcal F_r, \mathcal S_{ir} \right)
\end{equation}
denote bidder $i$'s forecast mean and variance of integrated variance for round $r$. Since $P_{T_r}$ is (approximately) known at $D_r$, we obtain
\begin{align}
  \E\!\left[\Pi_{ir} \mid \mathcal F_r, \mathcal S_{ir} \right]
    &= \alpha_{ir} + \beta_{ir} m^{\IV}_{ir} \sqrt{P_{T_r}} \, , \label{eq:profit-mean} \\
  \Var\!\left(\Pi_{ir} \mid \mathcal F_r, \mathcal S_{ir} \right)
    &= \beta_{ir}^2 v^{\IV}_{ir} P_{T_r} \, . \label{eq:profit-var}
\end{align}
We now introduce a key assumption: the conditional independence of bidders' valuation and risk aversion.

\begin{assump}[CIPV]\label{ass:cipv}
Conditional on the public information $\mathcal F_r$, bidders' valuations parameters \[\{(\alpha_{ir}, \beta_{ir}, \gamma_{ir}, m^{\IV}_{ir}, v^{\IV}_{ir})\}_{i=1}^N\] are independently drawn.
\end{assump}

\noindent Assumption~\ref{ass:cipv} formalizes the ``conditionally independent private values'' (CIPV) environment: public information $\mathcal F_r$ is common to all bidders, while private signals $\mathcal S_{ir}$ generate heterogeneous valuations through the parameters \[(\alpha_{ir}, \beta_{ir}, \gamma_{ir}, m^{\IV}_{ir}, v^{\IV}_{ir}),\] which reflects the reality: while there are public signals observable to everyone, such as recent prices and order flow on exchanges, each firm relies on one's proprietary models to forecast future volatility, and each firm's edge or alpha of finding optimal route or executing trades remain private.

\begin{assump}[Preferences]\label{ass:preferences}
Each bidder $i$ is risk-averse and evaluates random profits using a mean--variance functional. That is, there exists a risk-aversion parameter $\rho_{ir} > 0$ such that the certainty equivalent (CE) of any profit random variable $X$ in round $r$ is
\[
  \CE_{ir}(X)
  = \E[X] - \frac{\rho_{ir}}{2} \Var(X).
\]
\end{assump}

Under Assumption~\ref{ass:preferences}, bidder $i$'s valuation of the express lane in round $r$ is the certainty equivalent of $\Pi_{ir}$:
\begin{equation}\label{eq:valuation-def}
  v_{ir}
  \equiv \CE_{ir}(\Pi_{ir})
  = \E\!\left[\Pi_{ir} \mid \mathcal F_r, \mathcal S_{ir} \right]
    - \frac{\rho_{ir}}{2}
      \Var\!\left(\Pi_{ir} \mid \mathcal F_r, \mathcal S_{ir} \right).
\end{equation}
Substituting \eqref{eq:profit-mean} and \eqref{eq:profit-var} into
\eqref{eq:valuation-def} yields
\begin{align}
  v_{ir}
    &= \alpha_{ir}
       + \beta_{ir} m^{\IV}_{ir} \sqrt{P_{T_r}}
       - \frac{\rho_{ir}}{2} \beta_{ir}^2 v^{\IV}_{ir} P_{T_r} \nonumber \\
    &= \alpha_{ir}
       + \beta_{ir} m^{\IV}_{ir} \sqrt{P_{T_r}}
       - \gamma_{ir} v^{\IV}_{ir} P_{T_r},
       \label{eq:valuation}
\end{align}
where we define
\[
  \gamma_{ir} \equiv \frac{\rho_{ir}}{2} \beta_{ir}^2 > 0.
\]
Equation~\eqref{eq:valuation} makes explicit that higher expected volatility (integrated variance) increases valuations, while greater uncertainty about volatility lowers valuations through the risk-aversion term.

\subsection{Auction Equilibrium and Empirical Implications}

Given the valuation structure in \eqref{eq:valuation}, we now describe bidding behavior in the express lane auction. In each round $r$, bidders simultaneously submit sealed bids $\{b_{ir}\}_{i=1}^N$. The bidder with the highest bid wins the right to use the express lane during $[T_r, T_{r+1})$ and pays the second-highest bid, while all other bidders pay zero. The following proposition justifies the use of the submitted bid as the bidder's valuation.

\begin{prop}[Equilibrium Bidding]\label{prop:truthful}
Under Assumptions~\ref{ass:preferences} and~\ref{ass:cipv}, in each round $r$ of the express lane auction, it is a (weakly) dominant strategy for each bidder to bid her valuation:
\[
  b_{ir} = v_{ir} \quad \text{for all } i \in \{1,\dots,N\}.
\]
\end{prop}

Proposition~\ref{prop:truthful} is the standard truth-telling result for second-price auctions with private values; see, for example, \cite{krishna2009auction}. Combining Proposition~\ref{prop:truthful} with \eqref{eq:valuation}, we obtain the structural relationship
\begin{equation}\label{eq:structural-bid}
  b_{ir}
  = \alpha_{ir}
    + \beta_{ir} m^{\IV}_{ir} \sqrt{P_{T_r}}
    - \gamma_{ir} v^{\IV}_{ir} P_{T_r}.
\end{equation}

Equation~\eqref{eq:structural-bid} shows that, in our model, equilibrium bids are increasing in the bidder's forecast of integrated variance and decreasing in the forecast uncertainty about integrated variance, with both effects scaled by the current asset price. This link motivates our empirical specification.

In the data, we do not observe the latent objects $(\alpha_{ir}, \beta_{ir}, \gamma_{ir}, m^{\IV}_{ir}, v^{\IV}_{ir})$ directly. Instead, we approximate \eqref{eq:structural-bid} by a reduced-form model in which expected integrated variance and its uncertainty are replaced by observable proxies and bidder- or round-specific components are absorbed into fixed effects. Specifically, in the subsequent section we estimate regressions of the form
\begin{equation}\label{eq:reduced-form}
  b_{ir}
  = \theta_{0,i}
    + \theta_{1,i} \widehat{\E [\IV_r]} \sqrt{P_{T_r}}
    + \theta_{2,i} \widehat{\Var (\IV_r)} \, P_{T_r}
    + \varepsilon_{ir},
\end{equation}
where:
\begin{itemize}
  \item $\widehat{\E [\IV_r]}$ is a proxy of forecast mean $m^{\IV}_{ir}$,
  \item $\widehat{\Var (\IV_r)}$ is a proxy for the forecast variance
        $v^{\IV}_{ir}$, and
  \item $\varepsilon_{ir}$ is an error term.
\end{itemize}
Under \eqref{eq:structural-bid}, the theory predicts $\theta_{1,i} > 0$ and $\theta_{2,i} < 0$: bids should increase with expected volatility (more MEV to extract) and decrease with uncertainty about volatility (stronger risk discount). These sign restrictions form our main testable hypotheses.

%% file: Empirical_Analysis.tex
\section{Empirical Analysis}

\subsection{Data}
\paragraph*{Auction Records.} For the auction bids, we obtained the data directly from Arbitrum's database \cite{offchainlabs2025howtousetimeboost}. Our sample spans bids from May to October 2025. While multiple bidders participate, we focus on the two most dominant participants, \texttt{0x8c6f} and \texttt{0x95c0}, who together win over $90\%$ of rounds between May and October. 

\paragraph*{Measuring Volatility.} We proxy market-wide volatility using high-frequency ETH price data from Binance. As ETH is the most liquid crypto-asset on Arbitrum\footnote{We exclude stablecoins; see \cite{l2beat}.}, focusing on ETH serves as a low-friction approximation of overall market conditions. While price follows a stochastic process and thus the integrated variance is also essentially random, all we have is historical data of a single \emph{realized} path out of all possible outcomes. Therefore, we have to estimate $\E[\IV]$ and $\Var(\IV)$ from the given path of price. We adopt a standard approach, treating the realized variance (RV), the sum of squares of log returns $r_t$, as $\widehat{\E[\IV]}$, while estimating $\widehat{\Var(\IV)}$ with the following Newey-West estimator: \[
\widehat{\Var(\IV)} = T (\gamma_0 + 2\sum_{k=1}^L w_k \gamma_k),
\] where \[
\gamma_k \equiv \frac{1}{T} \sum_{t=1}^{T - k} (r^2_{t + k} - \overline{r^2})(r^2_t - \overline{r^2}) \quad \text{and} \quad \overline{r^2} \equiv \frac{1}{T} \sum_{t=1}^T r_t^2\, ,
\] for $k = 0, 1, 2, \cdots$, and $w_k \equiv 1 - \frac{k}{L + 1}$ is weight function. The standard practice of setting \(L \) is of \( L = \mathcal O(T^{1/3})\). In our setting, we set $T = 60$ and $L = 5$.

\paragraph*{Bidder Forecasts.} The theoretical model proposed in \cref{sec:theory} posits that bids depend on bidders' private forecasts of realized variance, which are unobservable. We therefore adopt the following assumption to bridge the theory with our ex-post volatility measures:

\begin{assump}[On-average Correct Forecast]
    Bidders are capable of \emph{on-average} correctly forecasting the mean and variance of forthcoming $\IV_r$ based on public and private information available at time $D_r$. That is, for \[
         \epsilon_{ir} \equiv m^{\IV}_{ir} - \E[\IV_r] \quad and \quad \tau_{ir} \equiv v^{\IV}_{ir} - \Var(\IV_r),
    \] \( \E \left[ \epsilon_{ir} | \mathcal F_r \right] = \E \left[ \tau_{ir} | \mathcal F_r\right] = 0\).\footnote{Note that this does not necessarily violate the CIPV framework. One can interpret this assumption that, based on public information, it is possible to, on average, correctly forecast the mean and variance of $\IV$. Then $\epsilon_{ir}$ and $\tau_{ir}$ being independent and private is sufficient for the CIPV setting.} 
\end{assump}
\noindent This errors-in-variables framework implies that our estimates of $\theta_1$ and $\theta_2$ will be subject to attenuation bias, which will bias them towards zero. Consequently, finding statistically significant coefficients despite this bias would provide a conservative and even stronger confirmation of our hypotheses.

\subsection{Estimation}
Our dependent variable, the observed bid $b_{ir}$, is left-censored at the reserve price $C = 0.001$ ETH. To account for the fat tails often observed in financial data, we employ a heteroskedastic Tobit model, assuming that the normalized residuals follow a Student's t distribution. Also, the bid unit is in ETH, while the bidder's valuation is in USD. Thus, we divide the bidder's valuation by the ETH price at each round.\footnote{The division on constant term is omitted, since the range of price (minimum $1.7$k and maximum $4.9$k USD) was negligible compared to that of the estimated mean and variance of \(\IV\).} Then, the model is defined as follows:
\begin{align}
    b_{ir} &= \max(C, v^*_{ir}) \label{eq:tobit_latent}\\
    v^*_{ir} &= \theta_0 + \theta_1 \frac{\widehat{\E[\IV]}}{\sqrt{P_r}} + \theta_2 \widehat{\Var(\IV)} + \varepsilon_{ir} \label{eq:tobit_mean}\\
    \log(\sigma_r) &= \gamma_0 + \gamma_1 \log\left(\frac{\widehat{\E[\IV]}}{\sqrt{P_r}}\right) + \gamma_2 
    \log\left( \widehat{\Var(\IV)} \right) \label{eq:tobit_scale} \\
    \frac{\varepsilon_{ir}}{\sigma_r} &\sim t(\nu)
\end{align}
where $v^*_{ir}$ is the latent valuation of bidder $i$ for round $r$, $b_{ir}$ is the observed bid, $\widehat{\E[\IV]}$ and $ \widehat{\Var(\IV)}$ are estimated mean and variance of integrated variance of given round, $\sigma_r$ is the conditional scale (standard deviation) of the latent valuation, modeled in log-linear form to ensure positivity, and $t(\nu)$ is the Student's t-distribution with $\nu$ degrees of freedom, which is also estimated. We also adjusted the scale of data ($\times 10^{-15}$ for bids, $\times 10^9$ for $\frac{\widehat{\E[\IV]}}{\sqrt{P_r}}$ and $\times 10^{12}$ for $\widehat{\Var(\IV)}$)  for numerical stability. Our hypotheses from the end of the previous section then translate to testing the coefficients in the mean model (Equation \ref{eq:tobit_mean}):
\begin{itemize}
    \item $\mathbf{H_1: \theta_1 > 0}$. A positive $\theta_1$ supports that higher expected volatility leads to higher valuations.
    \item $\mathbf{H_2: \theta_2 < 0}$. A negative $\theta_2$ supports the conclusion that higher uncertainty (forecast variance) leads to a larger risk-aversion discount and thus to lower valuations.
\end{itemize}

In addition to the linear specification derived from the mean-variance framework, we also estimate a log-log specification (where both bids and volatility measures are logged). While the linear model follows directly from our theoretical assumptions, the log-log specification offers strong empirical justifications. First, it systematically enforces nonnegative valuations, ruling out the theoretical possibility of negative bids implied by linear risk adjustments. Second, since the integrated variance and bids are numerically small values spanning multiple orders of magnitude, the log-log transformation mitigates the influence of extreme outliers and heteroskedasticity. Under these conditions, the log-log form can be interpreted as a first-order approximation of the equilibrium relationship, capturing potential nonlinearities in risk aversion (e.g., CRRA preferences) while serving as a strict robustness check for our main findings. 

\subsection{Results} 

\begin{figure}[htbp]
    \centering
    \includegraphics[width=0.9\linewidth]{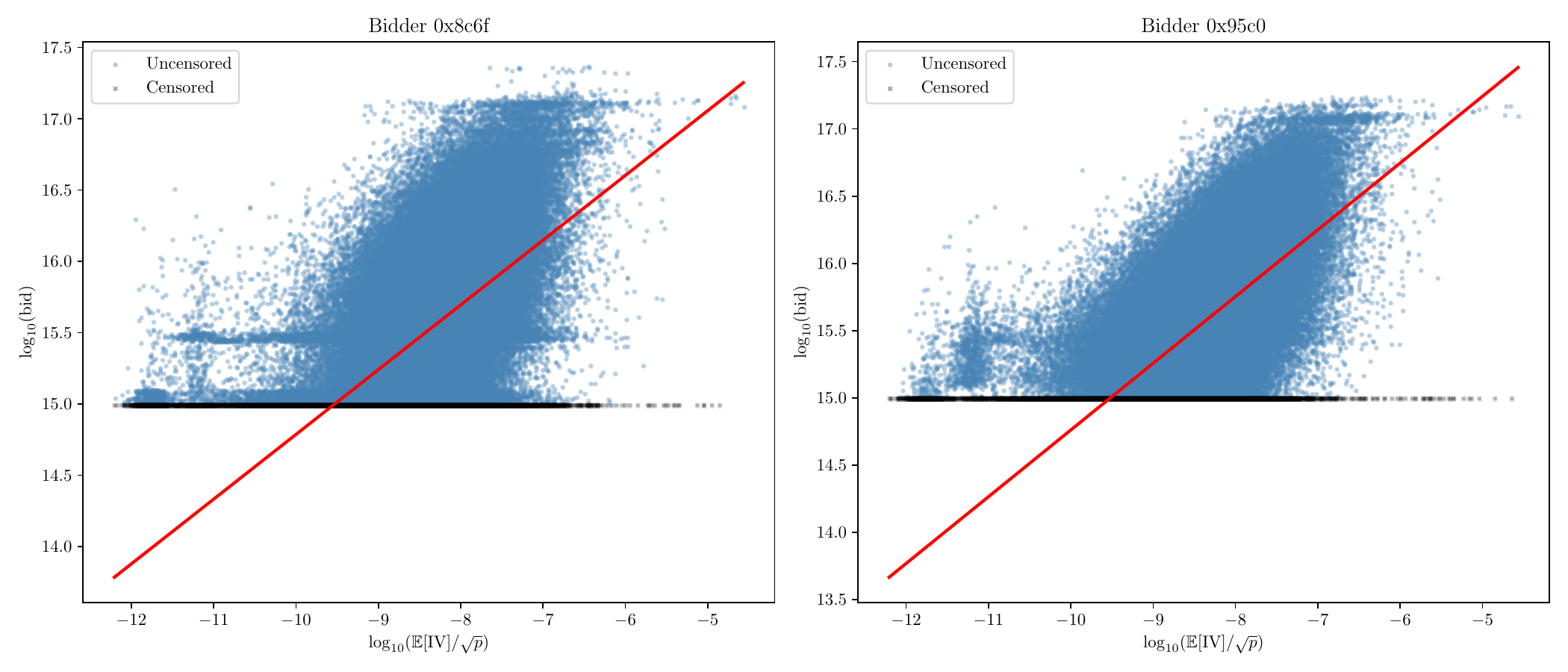}
    \caption{Scatter plot of each bidder's bids and price-adjusted integrated variance. Both are shown in log scale for visibility.}
    \label{fig:bid_vs_eiv}
\end{figure}

\Cref{tab:linear_nolag} shows the result of the estimates. For comparison, we also attach the result from the restricted model, which is identical to the original model except that the latent valuation depends only on $\widehat{\E[\IV]}$. This corresponds to the risk-neutral bidder's valuation. The first column corresponds to the restricted model, and the second to the previously introduced full model.

\begin{figure}[htbp]
    \centering
    \includegraphics[width=0.9\linewidth]{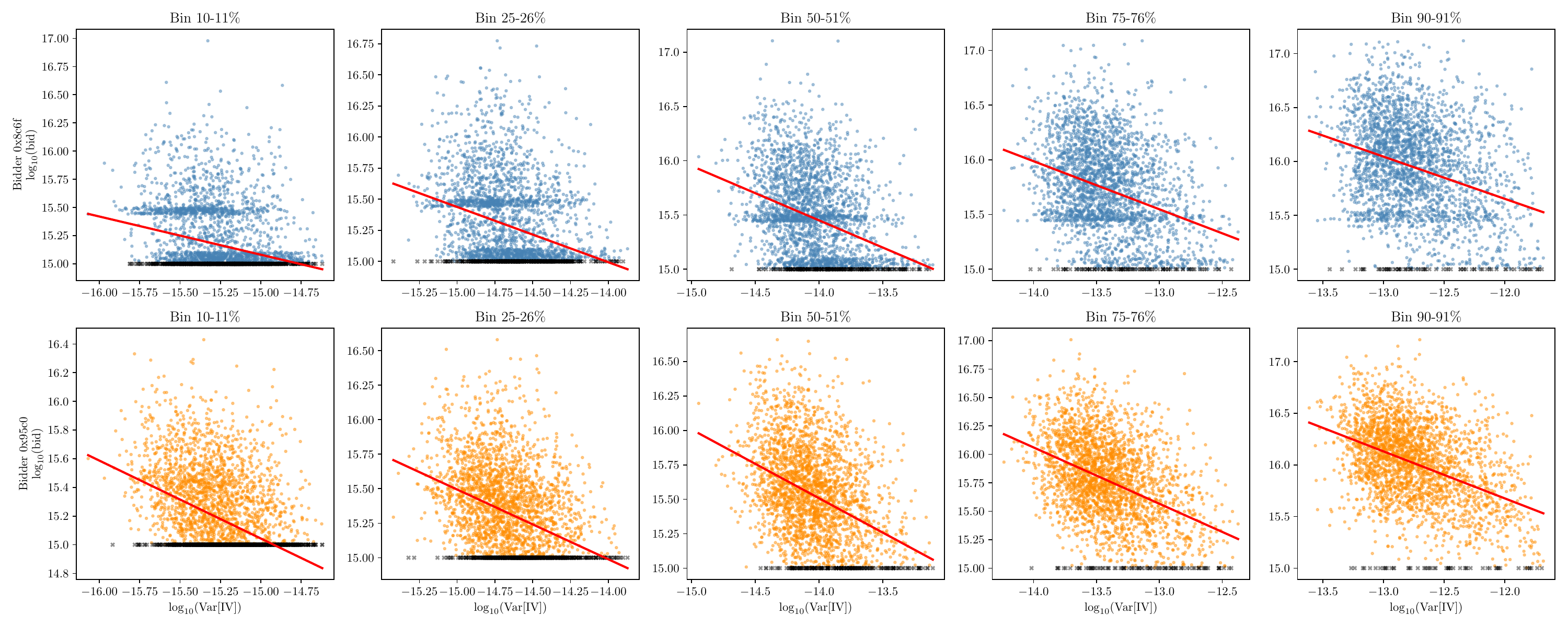}
    \caption{Scatter plots of each bidder's bids grouped by percentile of $\E[\IV]/\sqrt P$. A heteroskedastic tobit regression was done for each subsample. For visibility, the figure is in log scale.}
    \label{fig:bid_vs_variv}
\end{figure}

Across bidders, augmenting the specification with a variance term significantly improves the fit: LR tests are significant, and information criteria drop substantially. In the location equation, $\theta_1>0$ indicates that a higher expected IV is associated with higher bids, consistent with the view that the express-lane option is more valuable when arbitrage intensity rises. Conversely, $\theta_2<0$ on the standardized variance of the IV implies a risk discount: greater forecast uncertainty lowers valuations. Estimated tail parameters ($\nu\approx1.2$–$1.7$) confirm heavy-tailed residuals, justifying the Student-$t$ specification. \Cref{fig:bid_vs_eiv,fig:bid_vs_variv} provide comprehensive visualization of our result. 

Interestingly, the log-log variant of the reduced and full models yielded better results, although it lacks a theoretical justification. This may be due to the aforementioned scaling issues with the independent variables. The magnitude of the estimated mean and variance of IV spans across \(10^{-6}\) to \(10^{-3}\) and \(10^{-22}\) to \(10^{-12}\), respectively, and residuals on the samples with large estimated IV moments dominate in the process of MLE. See \cref{tab:log_nolag}.

\begin{table}[htbp]
\centering
\begin{tabular}{lcccc}
\toprule
 & \multicolumn{2}{c}{Bidder 0x8c6f} & \multicolumn{2}{c}{Bidder 0x95c0} \\
\cmidrule(lr){2-3} \cmidrule(lr){4-5}
Variable & Reduced & Full & Reduced & Full \\
\midrule
\multicolumn{5}{l}{\textit{Panel A: Location Coefficients}} \\
\addlinespace[0.5em]
Intercept ($\theta_0$) & 0.9878$^{***}$ & 0.9836$^{***}$ & 1.3691$^{***}$ & 1.1393$^{***}$ \\
 & (0.0011) & (0.0021) & (0.0061) & (0.0060) \\
E[IV]/$\sqrt{P}$ ($\theta_1$) & 0.3404$^{***}$ & 0.3472$^{***}$ & 0.3828$^{***}$ & 0.4421$^{***}$ \\
 & (0.0012) & (0.0014) & (0.0015) & (0.0017) \\
Var(IV) ($\theta_2$) & -- & -2.0792$^{***}$ & -- & -3.1878$^{***}$ \\
 & -- & (0.0340) & -- & (0.0460) \\
\addlinespace[0.5em]
\midrule
\multicolumn{5}{l}{\textit{Panel B: Scale Coefficients}} \\
\addlinespace[0.5em]
Intercept ($\gamma_0$) & -0.1690$^{***}$ & -2.2705$^{***}$ & 0.2718$^{***}$ & -1.4502$^{***}$ \\
 & (0.0052) & (0.0286) & (0.0041) & (0.0252) \\
$\log$(E[IV]/$\sqrt{P}$) ($\gamma_1$) & 0.5933$^{***}$ & 1.0895$^{***}$ & 0.4310$^{***}$ & 0.8572$^{***}$ \\
 & (0.0022) & (0.0072) & (0.0018) & (0.0064) \\
$\log$(Var(IV)) ($\gamma_2$) & -- & -0.3005$^{***}$ & -- & -0.2390$^{***}$ \\
 & -- & (0.0039) & -- & (0.0034) \\
\addlinespace[0.5em]
\midrule
\midrule
\multicolumn{5}{l}{\textit{Panel C: Model Fit Statistics}} \\
\addlinespace[0.5em]
Observations & 264959 & 264959 & 264959 & 264959 \\
Censored & 264959 & 264959 & 264959 & 264959 \\
Student $t$ $\nu$ & 1.23 & 1.30 & 1.55 & 1.70 \\
Log Likelihood & -729873.32 & -724012.17 & -724400.50 & -717011.02 \\
AIC & 1459756.64 & 1448038.34 & 1448810.99 & 1434036.03 \\
BIC & 1459809.08 & 1448111.75 & 1448863.43 & 1434109.44 \\
McFadden $R^2$ & 0.071 & 0.079 & 0.078 & 0.087 \\
\addlinespace[0.5em]
\midrule
\multicolumn{5}{l}{\textit{Panel D: Full vs Reduced Model Comparison}} \\
\addlinespace[0.5em]
$\Delta$ AIC (Full - Reduced) & \multicolumn{2}{c}{-11718.30} & \multicolumn{2}{c}{-14774.96} \\
$\Delta$ BIC (Full - Reduced) & \multicolumn{2}{c}{-11697.32} & \multicolumn{2}{c}{-14753.99} \\
LR Test $\chi^2$ (p-val) & \multicolumn{2}{c}{$11722.30^{***}$} & \multicolumn{2}{c}{$14778.96^{***}$} \\
\bottomrule
\end{tabular}
\caption{Heteroskedastic Tobit (Linear, t-dist Errors). $^{*}p<0.05$, $^{**}p<0.01$, $^{***}p<0.001$.}
\label{tab:linear_nolag}
\end{table}

\paragraph*{Robustness.}
We assess the robustness of the estimated relationships across different market conditions. To this end, we partition the dataset along both the temporal and regime dimensions. Specifically, the full sample is divided into (i) monthly subsamples from May through October and (ii) two volatility regimes (low and high, with median as threshold) delineated by the cross-sectional median of volatility. The regression model is then re-estimated separately for each subsample, and the corresponding estimates are reported in \cref{tab:tobit_subsample}.

Although the magnitude of the estimated coefficients varies across subsamples, the sign pattern is preserved: the coefficient on the variance of variance remains negative (interpreted as a discount) in all cases. A notable exception arises in the low-volatility regime, where the estimate of the coefficient on \(\theta_2\) is disproportionately large. While we do not have a definitive explanation for this outcome, we conjecture that it is attributable to a scaling issue. In particular, \(\widehat{\Var(\IV)}\) is approximately of the same order of magnitude as \(\widehat{\E[\IV]}^2\), with its minimum value on the order of \(10^{-22}\), and such extreme scaling potentially lead to estimation bias and numerical instability in some cases.

To further verify our hypothesis without relying on the rational-expectations, i.e., the on-average correct forecasts assumption, we also ran a regression using lagged IV moment estimators, the estimated mean and variance of IV from the previous round, which are available to bidders at the time of bidding. For the result, see \cref{tab:linear_lagged,tab:log_lagged}. We found that not only does the discount remain persistent, but the overall goodness of fit also improves significantly for both linear and log-transformed models, suggesting the possibility of bidders relying on a simple AR(1) model instead of trying to precisely forecast the volatility distribution.

\subsection{Discussion}
The empirical findings reveal a statistically significant discount attributed to the inherent uncertainty (i.e., risk) associated with forecasting future volatility and the generated arbitrage revenue. 
This suggests that the efficiency of the ELA may diminish during periods of high market turbulence, not just because volatility is high, but because it becomes less predictable. This discount leads to lower bids relative to bidders' expected value of CEX-DEX arbitrage profit and, consequently, to a lower MEV capture rate. Further refinement of the ELA mechanism is required to mitigate this observed inefficiency. 

Given the challenges of accurately forecasting short-term volatility due to its low signal-to-noise ratio and non-stationarity, extending the round length to a longer duration, such as hourly or even daily, may be a more robust design. While longer horizons introduce other challenges, such as increased capital requirements, they may filter high-frequency noise and improve the signal-to-noise ratio of volatility forecasts. We also note that, instead of a seemingly arbitrary $0.001$ ETH, the reserve price should be set strategically to maximize expected revenue, balancing the guaranteed lower bound on payment with the probability of failing to attract any bidders.

Alternatively, a subscription-based model could be considered for simultaneously offering the express lane to multiple market participants. This approach has established precedents within traditional financial (TradFi) exchanges \cite{budish2024theory}, and recently launched perpetual exchanges such as Lighter and Paradex are adopting it as well\footnote{See \url{https://apidocs.lighter.xyz/docs/account-types} and \url{https://docs.paradex.trade/trading/trader-profiles\#fee-structure}.}.

\paragraph*{Limitations.}
Although the data support our hypotheses, our model has several limitations. Firstly, our model presents an oversimplified representation. Factors such as Total Value Locked (TVL) and liquidity near the current price significantly influence bidders' revenue; however, these elements are not incorporated into our model, which may introduce bias into our estimations. Secondly, data noise poses a challenge. Binance candles, generated from the last traded price, are susceptible to sudden price spikes from large market orders, thereby becoming decoupled from the true price. The use of future values as a proxy for the forecast also causes attenuation bias. We also remark that a significant portion of rounds being censored in samples and subsamples further weakens our claims, too.

%% file: Conclusion.tex
\section{Conclusion}
In this study, we hypothesized that Timeboost exhibits a lower MEV capture rate under volatile market conditions and empirically tested this hypothesis. Our methodology uses realized values to construct a noisy proxy for an unobserved variable that captures bidders' forecasts of future volatility. We developed a bidder valuation model and proposed using the coefficient on the variance of integrated variance ($\Var(\IV)$) as the risk premium for the uncertainty in the value of access to the express lane (EL). 
Subsequently, we compared the regression results of our model with those of a baseline model that excluded a risk premium, demonstrating improvements in fit and a reduction in downward bias in the restricted model. We further tested our hypothesis under various conditions to check the robustness of the results.

This study shows that variance risk premiums are statistically significant and negative. Specifically, it demonstrates that the full model provides a better explanation of the findings compared to the restricted model. As expected, the restricted model yields a downward-biased slope relative to the full model, reinforcing the overall argument. We further tested our hypothesis under various conditions, checking the results with subsamples, using alternative measures of volatility, dropping the rational-expectations assumption, and using forecasted rather than realized values to assess robustness.

Based on the results, we provided an outline to improve the current Timeboost auction implementation and increase the sequencer's revenue. Specifically, we offered two suggestions: a more advanced method for setting the reserve price and extending the round length, which helps bidders better estimate future (averaged) volatility, leading to less discounting. We also briefly mentioned the option of switching to a subscription model, which is already popular in traditional finance (TradFi).

Our methodology exhibits several notable limitations. Primarily, the model's simplicity results in the omission of numerous real-world complexities. For example, the model assumes that the liquidity available around the CEX price remains constant throughout the analysis period and restricts its scope to ETH as the only volatile asset within AMMs. Furthermore, reliance on the CIPV framework may not fully capture the intricacies of actual market dynamics. The use of future realized variance and Newey-West estimators, which are inherently unknown at the time of bidding, as noisy proxies for bidders’ forecasts, introduces attenuation bias. Finally, the substantial rate of data censorship poses significant challenges to the robustness and validity of the inferences drawn, despite the use of tobit regression.

Despite of aforementioned limitations, this study is one of the first empirical investigations of Timeboost, examining the determinants of bidders' valuations and the efficacy of the express lane auctions. As such, it provides a foundational reference for subsequent research that employs more advanced methodologies and incorporates realistic assumptions. Ultimately, this study aims to enhance the understanding of time-based sequencing mechanisms within auction frameworks.